\documentstyle[11pt,iau185,twoside,epsf]{article}

\begin{document}
\title{Resonant Excitation of Nonradial Modes in RR Lyrae 
Stars}
\author{R. Nowakowski}
\affil{Copernicus Astronomical Center, Warsaw, Poland}

\begin{abstract}
A nonlinear development of radial pulsation instability to 
a resonant excitation
of nonradial modes is studied. The theory covers the cases 
of axisymmetric
$(m=0)$ modes as well as $(m,-m)$ pairs. Adopting a 
simplified treatment of the
mode coupling, it is found that multimode pulsation with 
constant amplitudes
is a highly probable solution.
An observable consequence of the $m=0$ mode excitation is 
randomness of pulsation
amplitude. The case of an $\ell=1$ mode is the most 
important because of a small
averaging effect and a large excitation probability.
A significant amplitude and phase modulation is predicted 
in the case of
excitation of an $\ell=1,m=\pm1$ pair. This may explain 
Blazhko-type modulation
in RR~Lyrae stars. If this model is correct, the modulation 
period is determined
by the rotation rate and the Brunt-V\"ais\"al\"a frequency 
in the deepest part
of the radiative envelope.
\end{abstract}

\keywords{Stars: oscillations, Stars: variables: RR Lyrae}

\section{Introduction}

One of the most intriguing questions concerning RR~Lyrae 
stars is the nature of
the long-term modulations that characterize about 20--30\% 
of RRab and 2--3\% of
RRc stars (e.g., Moskalik \& Poretti, these proceedings; Kov\'acs, 2000).
The modulations of the oscillation amplitudes and phases 
manifest themselves in
the Fourier spectra as secondary peaks with frequencies 
close to the main
pulsation frequency. There seems to exist two 
distinguishable cases: one
secondary peak with frequency either higher or lower than 
the main frequency and
a pair of side peaks which, together with the main peak, 
form an equidistant
triplet.

There are two models proposed to explain the modulations: 
a magnetic oblique
rotator (Shibahashi, 2000) and the 1:1 resonance model 
(Van~Hoolst et al.,
1998). In this paper we develop the resonant model. 

\section{Nonresonant properties of oscillation modes}
As was shown by Van~Hoolst et al. (1998) and Dziembowski 
\& Cassisi (1999),
there always exists a very dense spectrum of low-$\ell$ 
nonradial modes in the
frequency range of the lowest radial overtones in models of 
RR~Lyrae stars. The
most unstable modes have frequencies very close to the 
fundamental as well as
the first overtone frequencies. Among the low degree modes 
in the closest
vicinity of the radial modes, the $\ell=1$ modes are the 
most unstable and thus
the most important dynamically.

Small frequency differences between the radial and the 
closest nonradial mode
suggest that 1:1 and similar resonances should be taken 
into account in the
study of the oscillations. Moreover, this frequency 
difference, which is an
important parameter in resonant dynamics, is such a 
sensitive function of the
stellar parameters that it is justified to treat it as a 
random quantity.

In the simplest nonlinear theory of stellar pulsations 
without resonances,
sufficiently high amplitude of one mode may stabilize all 
other linearly
unstable modes so that they are not present in 
oscillations. This simple
saturation mechanism leads to monomode pulsations which are 
typical for RR~Lyrae
stars. 
\section{Resonant coupling and instability of monomode 
pulsations}
One of the conditions for the resonant coupling between 
different modes may
be written as $\sum_i m_i=0$, where $m_i$ are azimuthal 
numbers of the
interacting modes. For the most important case of $\ell=1$ 
it means two
possibilities:
\begin{itemize}
\item coupling of the radial mode with an $m=0$ nonradial 
mode,
\item coupling of the radial mode with an $m=1,m=-1$ pair.
\end{itemize}
It turns out that the two cases are dynamically equivalent, 
i.e., the amplitude
equations describing the mode amplitudes are in both cases 
the same with the
same coefficients and in the latter case 
$A_{1,1}=A_{1,-1}$.

Radial monomode pulsations become resonantly unstable when
\begin{equation}
|\Delta\omega|<\sqrt{|R||A_{0,0}|^2-\kappa^2_{l,N}}
\end{equation}
(Van Hoolst et al., 1998), where $\kappa_{l,N}$ is a 
nonlinear growth rate of
the nonradial mode, $R$ is a resonant coupling coefficient, 
$A_{0,0}$ is a
radial mode amplitude, and the detuning parameter is given 
by
\begin{displaymath}
\Delta\omega=\left\{
\begin{array}{ll}
\omega_{1,0}-\omega_{0,0}\ &{\rm for}\ m=0\\
(\omega_{1,1}+\omega_{1,-1})/2-\omega_{0,0}\ &{\rm for}\ 
m=\pm1\ 
{\rm pair.}
\end{array}\right.
\end{displaymath}

\section{Finite amplitude development of resonant 
instability}

In order to simplify the problem we made some assumptions. 
The most important
are the use of adiabatic resonant coupling 
coefficients and the neglect of
nonresonant nonlinear frequency changes. The former 
assumption was used by
Van~Hoolst et al. (1998) and Dziembowski \& Cassisi (1999). 
The latter
assumption seems to be justified because the nonlinear 
frequency change has the
same sign for both modes and it does not change the 
frequency difference
significantly.

We first studied double-mode stationary solutions of 
amplitude equations
together with their stability with respect to small 
perturbations. The
parameters of the equations were calculated for the range 
of stellar models by
Dziembowski \& Cassisi (1999). A typical solution is 
presented in Fig.~1. All
interesting pairs of modes in all models give qualitatively 
the same solutions.

\begin{figure}[t]
\begin{center}
\mbox{\epsfxsize=0.5\textwidth\epsfysize=0.45\textwidth\epsfbox{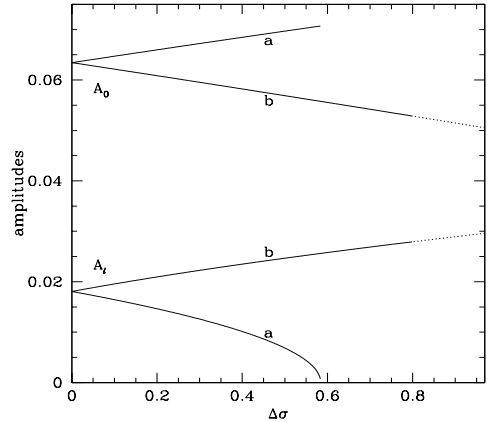}}
\caption{Double-mode stationary solutions for the radial 
fundamental mode (thick
line) and the nearest $\ell=1,m=0$ nonradial mode (thin 
line) for a chosen RR~Lyrae
model. The dimensionless detuning parameter $\Delta\sigma$ 
is defined as
$\Delta\omega/\kappa_{0,0}$. The problem is symmetric with 
respect to the change
of the sign of the detuning parameter and the presented 
range 
of this parameter
corresponds to the maximum value given by the density of 
the nonradial modes
spectrum. Dotted lines indicate unstable solutions. For 
$\Delta\sigma$ smaller
than approximately $0.6$ there is a pair of solutions 
denoted by $a$ and $b$.}
\end{center}
\end{figure}
The $a$-branch solution exists if the detuning parameter is 
smaller than the
critical value given by Eq.\,(1) (about $0.6$ in the case 
presented in Fig.\,1)
and it is always stable. If the detuning parameter is 
higher than the previously
mentioned critical value the monomode solution (not shown 
in Fig.~1) is stable,
according to Eq.\,(1). This means that there always exists 
at least one stable
fixed-point solution, either monomode (large 
$\Delta\omega$) or double-mode
(small $\Delta\omega$).

Direct time integration of the amplitude equations shows 
that the time-dependent
solutions always converge to one of the stable fixed-point 
solutions.

Multimode fixed-point solutions of resonant amplitude 
equations are
characterized by the {\em phase-lock} phenomenon (for the 
general discussion see
Buchler et al., 1997). In our cases it means:
\begin{itemize}
\item for interaction with an $m=0$ nonradial mode, the 
two frequencies are equal,
\item for interaction with an $m=\pm1$ pair, the three 
frequencies are
equidistant.
\end{itemize}

\section{Observational consequences}

When resonant interaction leads to excitation of an $m=0$ 
nonradial mode, an
observer would see only one frequency due to the phase-lock 
phenomenon. The
only effect of the presence of the nonradial mode is the 
aspect-dependence and
thus the randomness of the observed pulsation amplitude.

The excitation of the $m=\pm1$ pair leads to the observed 
triplet-type
modulation, where the side peaks in the spectrum are 
rotationally split $m=\pm1$ modes. A simple estimate yields 
the splitting to 
be equal to
$\bar\Omega/2$, where $\bar\Omega$ is the mean rotation 
rate weighted with the
Brunt-V\"ais\"al\"a frequency which is strongly peaked in 
the deepest part of
the radiative envelope. Thus, in this model the Blazhko 
period is the measure of
the rotation in this region. 
\section{Conclusions and future work}

The resonant model presented in this paper predicts many 
observed features of
RR~Lyrae stars: random occurrence and ranges of the 
modulations as well as lower
occurence rate of Blazhko stars among RRc than among RRab 
stars.

However, this model has some problems, too. The most 
important one is that it
does not predict the strong asymmetry of the side-peak 
amplitudes (Kov\'acs,
these proceedings). Moreover, it does not explain features 
which have been
recently observed: larger frequency splittings among RRc 
than among RRab
stars (Moskalik \& Poretti, these proceedings) and the 
change of the
Blazhko-period (Smith et al., these proceedings).

To understand the nature of the long-term modulations of 
RR~Lyrae stars more
efforts should be made both theoretically and 
observationally. The most
promising are attempts to find the nonradial mode 
characteristics in
line-profile variations of \object{RR\,Lyrae} itself 
(Kolenberg et al., these
proceedings). Precise spectroscopic observations of other 
Blazhko stars would
also be of great importance.

Our model also needs some improvements. The most important 
one is to take into
account previously neglected coupling coefficients and more 
interacting modes,
in particular the whole $\ell=1$ triplet.\\[0.5cm]

{\em This paper summarizes the results given by
Nowakowski \& Dziembowski (2001)} 

\acknowledgements We would like to thank the organizers for 
the financial
support. We are also greatful to Wojtek Dziembowski and 
Geza Kov\'acs for their
useful suggestions given during the preparation of our talk 
and this paper.

\end{document}